\documentstyle[aps,preprint,prc,12pt,epsf]{revtex}

\newcommand{\be}{\begin{equation}}
\newcommand{\ee}{\end{equation}}
\newcommand{\ba}{\begin{eqnarray}}
\newcommand{\ea}{\end{eqnarray}}

\newcommand{\ave}[1]{\langle {#1} \rangle}
\tightenlines

\begin{document}

\title{
Moments of event observable distributions and many-body correlations
}

\author{
M. Belkacem$^1$\footnote{E-mail: belkacem@th.physik.uni-frankfurt.de}, 
Z. Aouissat$^2$, M. Bleicher$^1$, H. St\"{o}cker$^1$ and 
W. Greiner$^1$
}

\address{
$^1$ Institut f\"{u}r Theoretische Physik, J. W. Goethe-Universit\"{a}t, \\
D-60054 Frankfurt am Main, Germany \\
$^2$ Institut f\"{u}r Kernphysik, Technische Hochschule 
Darmstadt, Schlo{\ss}gartenstra{\ss}e 9, \\
D-64289 Darmstadt, Germany
}

%\date{ \today }
\maketitle

\begin{abstract}

We investigate event-by-event fluctuations for ensembles with non-fixed
multiplicity. Moments of event observable distributions, like total energy
distribution, total transverse momentum distribution, etc, are shown to be 
related to the multi-body correlations present in the system. For classical
systems, these
moments reduce in the absence of any correlations to the moments of
particle inclusive momentum distribution. 
As a consequence, a zero value for the 
recently introduced $\Phi$-variable 
is shown to indicate the vanishing of two-body
correlations from one part, and of correlations between multiplicity and
momentum distributions from the other part.
It is often misunderstood as a measure of the degree of
equilibration in the system.

\end{abstract}

{
\vskip 2\baselineskip
%{\bf PACS : } 
}

% body of the paper
\newpage

%\narrowtext
%\begin{multicols}{2}

One of the main goals of relativistic heavy-ion collisions is the study of
hadronic matter under extreme conditions of temperature and density. This
offers the unique opportunity to investigate the possible phase transition of
hadronic matter to quark-gluon plasma. As is well known from statistical
mechanics \cite{huang,balescu}, large fluctuations may occur at such a phase
transition. These fluctuations are maximum at the critical point where large
portions of the system become strongly correlated. Moreover, the investigation 
of these
fluctuations is now possible with the advent of large acceptance detectors
which allow for the first time an event-by-event analysis of the data. 
Already, event-by-event fluctuations of the transverse momentum distributions 
have been proposed to provide information about the heat capacity 
\cite{stod,shur,blei1,step}, or about a possible equilibration of the system
\cite{gazd1,gazd2,mrow1,mrow2,blei2,gazd3}. On the experimental side, 
first preliminary
result of the NA49 collaboration seem to indicate the absence of any
non-statistical fluctuations in the mean transverse momentum distribution for
Pb-Pb collisions at 160 AGeV \cite{roland}. 

In this letter, we propose a general method to investigate the presence of 
many-body
correlations and of non-statistical fluctuations in momentum distributions of
multiparticle events. The method applies to
all p-p, p-A, or A-A collisions. The recently introduced $\Phi$-variable 
\cite{gazd1} appears to be one of the moments 
proposed to give
evidence for the presence or no of these correlations.

Consider the global observable defined for each event by:
\be
Z = \sum_{i=1}^{N} y(\vec{p_i}),
\label{eq1}
\ee
where $N$ indicates the multiplicity of the event considered and $y(\vec{p_i})$
is any function which depends of the momentum of particle $i$ in the event.
This quantity could be for instance the energy, the transverse momentum, etc. We
are interested by the fluctuations of this global observable from event to event
and in particular by the moments
\be
\ave{Z^{k}} = \frac{1}{M} \sum_{j=1}^{M}\left[\sum_{i=1}^{N_j}
y(\vec{p_i})\right]^{k},
\label{eq2}
\ee
where $M$ is the total number of events and $N_j$ indicates the multiplicity of
event $j$. 

Consider now the N-body distribution function 
$f_N(N,\vec{p_1},\cdot\cdot\cdot,\vec{p_N})$ which gives the probability for 
a system of $N$ particles where particle $1$
has a momentum $\vec{p_1}$, particle 2 a momentum $\vec{p_2}$, and so on. 
Since
we want to describe systems with different multiplicity, the distribution
function $f_N(N,\vec{p_1},\cdot\cdot\cdot,\vec{p_N})$ may as well depend on 
the multiplicity $N$. It is defined such that 
\be
\int d\vec{p_1} \cdot\cdot\cdot d\vec{p_N} ~
f_N(N,\vec{p_1},\cdot\cdot\cdot,\vec{p_N}) = P(N),
\label{eq3}
\ee
where $P(N)$, the probability of finding the system with exactly $N$
particles regardless of their momenta, is normalized according to :
\be
\sum_{N=0}^{\infty} P(N) = 1.
\label{eq4}
\ee

The reduced s-body distribution functions ($s<N$) for a system of
indistinguishable particles is given by \cite{huang,balescu}:

%\end{multicols}
%\widetext
%\llinea

\be
f_s(N,\vec{p_1},\cdot\cdot\cdot,\vec{p_s}) = \frac{N!}{(N-s)!} 
\int d\vec{p_{s+1}} \cdot\cdot\cdot d\vec{p_N} ~
f_N(N,\vec{p_1},\cdot\cdot\cdot,\vec{p_N}).
\label{eq5}
\ee

%In particular,
%\ba
%&&f_3(N,\vec{p_1},\vec{p_2},\vec{p_3}) = N(N-1)(N-2) \int d\vec{p_{4}}
%\cdot\cdot\cdot d\vec{p_N} ~ f_N(N,\vec{p_1},\cdot\cdot\cdot,\vec{p_N}); \\
%&&f_2(N,\vec{p_1},\vec{p_2})           = N(N-1) \int d\vec{p_{3}}
%\cdot\cdot\cdot d\vec{p_N} ~ f_N(N,\vec{p_1},\cdot\cdot\cdot,\vec{p_N}); \\
%&&f_1(N,\vec{p_1})                     = N \int d\vec{p_{2}}
%\cdot\cdot\cdot d\vec{p_N} ~ f_N(N,\vec{p_1},\cdot\cdot\cdot,\vec{p_N}).
%\ea

%\rlinea
%\narrowtext
%\begin{multicols}{2}

From the above definitions, we have:

\be
\int d\vec{p_1} \cdot\cdot\cdot d\vec{p_s} ~ 
f_s(N,\vec{p_1},\cdot\cdot\cdot,\vec{p_s}) = \frac{N!}{(N-s)!} P(N).
\label{eq6}
\ee

After these definitions, the moments of the event variable 
$Z$ are defined as:

%\end{multicols}
%\widetext
%\llinea

\be
\ave{Z^k} = \sum_{N=0}^{\infty} \int d\vec{p_1} \cdot\cdot\cdot d\vec{p_N} ~ 
\left[\sum_{i=1}^{N} y(\vec{p_i})\right]^{k} 
f_N(N,\vec{p_1},\cdot\cdot\cdot,\vec{p_N}),
\label{eq7}
\ee
and in particular,

\be
\ave{Z} = \sum_{N=0}^{\infty} \int d\vec{p_1} ~ 
y(\vec{p_1}) ~ f_1(N,\vec{p_1});
\label{eq8}
\ee
\be
\ave{Z^2} = \sum_{N=0}^{\infty} \left[ \int d\vec{p_1}~  y^2(\vec{p_1}) ~ 
f_1(N,\vec{p_1}) + \int d\vec{p_1}d\vec{p_2} ~ 
y(\vec{p_1})y(\vec{p_2}) ~ f_2(N,\vec{p_1},\vec{p_2}) \right];
\label{eq9}
\ee
\ba
\nonumber
&&\ave{Z^3} = \sum_{N=0}^{\infty} \left[ \int d\vec{p_1} ~ y^3(\vec{p_1}) ~ 
f_1(N,\vec{p_1}) + \int d\vec{p_1}d\vec{p_2} ~ 
y^2(\vec{p_1})y(\vec{p_2}) ~ f_2(N,\vec{p_1},\vec{p_2}) \right.\\
&&~~~~~~~~~~~~\left. + \int d\vec{p_1}d\vec{p_2}d\vec{p_3} ~ 
y(\vec{p_1})y(\vec{p_2})y(\vec{p_3}) ~ 
f_3(N,\vec{p_1},\vec{p_2},\vec{p_3}) \right],
\label{eq10}
\ea

%\rlinea
%\narrowtext
%\begin{multicols}{2}

\noindent
and so on for the higher moments. 
One sees that the fluctuations of the event observable $Z$ are
related to the higher n-body correlations; the second moment is related to
2-body correlations, the third moment to 2- and 3-body correlations and so on.
Note also that these moments are related to the possible
correlation of the multiplicity of particles to the s-body momentum 
distributions $f_s(N,\vec{p_1},\cdot\cdot\cdot,\vec{p_s})$\footnote{It seems
that what makes 
the s-body momentum distributions possibly depend on the
particle multiplicity are precisely the multi-body correlations. In the 
absence of correlations, the many-body distribution functions
consist of a product of one-body distribution functions. In this case, 
every particle in the system does not feel the presence of the other particles.
Its distribution function should not then depend on whether there is only
one particle or many of them, hence it should not depend on the multiplicity 
of particles in the
system.}. This result is similar to that obtained in \cite{bialas}.

Let us now answer the following question: how do the above defined moments of 
the
event variable $Z$ reduce in the absence of any correlation? If no correlations
are present in the system, the s-body distribution functions, consistent 
with Eqs.(\ref{eq3}-\ref{eq6}), read:
\be
f_s(N,\vec{p_1},\cdot\cdot\cdot,\vec{p_s}) = \frac{N!}{(N-s)!} P(N) ~ 
\widetilde{f}_1(\vec{p_1}) \cdot\cdot\cdot \widetilde{f}_1(\vec{p_s}),
\label{eq11}
\ee
with 
\be
\nonumber
\int d\vec{p} ~ \widetilde{f}_1(\vec{p}) = 1.
\ee
Eqs.(\ref{eq8}-\ref{eq10}) reduce then to

%\end{multicols}
%\widetext
%\llinea

\be
\ave{Z} = \ave{N} \int d\vec{p} ~ y(\vec{p}) ~ \widetilde{f}_1(\vec{p});
\label{eq12}
\ee
\be
\ave{Z^2} = \ave{N} \int d\vec{p} ~ y^2(\vec{p}) ~ \widetilde{f}_1(\vec{p})
     + \left(\ave{N^2} - \ave{N}\right) \left[\int d\vec{p} ~ y(\vec{p}) ~ 
       \widetilde{f}_1(\vec{p})\right]^2;
\label{eq13}
\ee
\ba
\nonumber
&&\ave{Z^3} = \ave{N} \int d\vec{p} ~ y^3(\vec{p}) ~ \widetilde{f}_1(\vec{p})
     + \left(\ave{N^2} - \ave{N}\right) \left[\int d\vec{p} ~ y^2(\vec{p}) ~ 
       \widetilde{f}_1(\vec{p})\right] \left[\int d\vec{p} ~ y(\vec{p}) ~ 
       \widetilde{f}_1(\vec{p})\right] \\
&&~~~~~~~~~+ \left(\ave{N^3} -3\ave{N^2} +2\ave{N}\right)\left[
       \int d\vec{p} ~ y(\vec{p}) ~ \widetilde{f}_1(\vec{p})\right]^3,
\label{eq14}
\ea

%\rlinea
%\narrowtext
%\begin{multicols}{2}

\noindent
where
\be
\nonumber
\ave{N^k} = \sum_{N=0}^{\infty} N^k ~ P(N).
\ee
\begin{enumerate}
\item
Note that in the absence of correlations, the
distribution function $\widetilde{f}_1(\vec{p})$ coincides with the
inclusive one-particle distribution function. Indeed, the inclusive
one-particle distribution function is defined\footnote{The inclusive
particle distribution function is defined such as the average value of a
given particle observable $O$ is given by: $\ave{O} = \frac{1}{N_{tot}} 
\sum_{i=1}^{N_{tot}}
O(\vec{p_i}) = \frac{1}{N_{tot}} \sum_{i=1}^{M} \sum_{j=1}^{N_i} 
O(\vec{p_{i,j}})$
where $M$ is the number of events and $\vec{p_{i,j}}$ is the momentum of
particle $j$ in event $i$. $N_{tot}$ is the total number of particles in 
all events; it is given by $N_{tot} = \sum_{i=1}^{M} N_i = M\ave{N}$ with 
$\ave{N} = \frac{1}{M} \sum_{i=1}^{M} N_i$. One obtains then $\ave{O} =
\frac{M}{N_{tot}} \frac{1}{M}\sum_{i=1}^{M} \sum_{j=1}^{N_i} O(\vec{p_{i,j}}) = 
\frac{1}{\ave{N}} \sum_{N=0}^{\infty} \int d\vec{p} ~ O(\vec{p}) ~ 
f_1(N,\vec{p_1})$ (see Eq.(\ref{eq8})).} as:
\be
f^{incl}(\vec{p}) = \frac{1}{\ave{N}} \sum_{N=0}^{\infty} f_1(N,\vec{p}),
\label{eq18}
\ee  
which reduces in the absence of correlations to
\be
f^{incl}(\vec{p}) = \frac{1}{\ave{N}} \sum_{N=0}^{\infty} N P(N) ~ 
\widetilde{f}_1(\vec{p}) \equiv \widetilde{f}_1(\vec{p}).
\label{eq19}
\ee 

\item
Note also that by a judicious choice of the particle variable $y(\vec{p})$ 
in such a way that in the absence of correlations, the mean value of 
$y(\vec{p})$ vanishes,
\be
\int d\vec{p} ~ y(\vec{p}) ~ \widetilde{f}_1(\vec{p}) = 0
\ee
the moments of the event observable $Z$ will be exactly proportional to the
moments of the particle observable $y(\vec{p})$.
A good
choice of the particle variable $y(\vec{p})$ is:
\be
y(\vec{p}) = x(\vec{p}) - \bar{x}
\label{eq23}
\ee
where $x(\vec{p})$ is any function of momentum $\vec{p}$ (kinetic-,
transverse-energy, transverse momentum, ... etc) and 
\be
\bar{x} = \int d\vec{p} ~ x(\vec{p}) ~ f^{incl}(\vec{p})
\label{eq24}
\ee
Note that with this choice, and according to Eqs.(\ref{eq8},\ref{eq18}), 
the average value of $Z$ is always zero, even in the presence of strong
correlations. 
\end{enumerate}

%
%\be
%\ave{Z} = \ave{N} \int d\vec{p} ~ y(\vec{p}) ~ \widetilde{f}_1(\vec{p}) 
%\equiv 0
%\label{eq15}
%\ee
%\be
%\ave{Z^2} = \ave{N} \int d\vec{p} ~ y^2(\vec{p}) ~ \widetilde{f}_1(\vec{p})
%\label{eq16}
%\ee
%\be
%\ave{Z^3} = \ave{N} \int d\vec{p} ~ y^3(\vec{p}) ~ \widetilde{f}_1(\vec{p})
%\label{eq17}
%\ee

It appears then that, in the absence of any correlations in the system and by 
a judicious 
choice of the particle variable $y(\vec{p})$, 
the moments of the event variable $Z$ are exactly proportional 
to the inclusive moments of the particle variable $y(\vec{p})$:
\be
\ave{Z} = \ave{N} \int d\vec{p} ~ y(\vec{p}) ~ f^{incl}(\vec{p}) \equiv 0 
\label{eq20}
\ee
\be
\ave{Z^2} = \ave{N} \int d\vec{p} ~ y^2(\vec{p}) ~ f^{incl}(\vec{p}) 
\label{eq21}
\ee
\be
\ave{Z^3} = \ave{N} \int d\vec{p} ~ y^3(\vec{p}) ~ f^{incl}(\vec{p})
\label{eq22}
\ee
The proportionality factor is the average number of particles $\ave{N}$. If
multi-body correlations are present in the system, Eqs.(\ref{eq21},\ref{eq22}) 
do not
hold any longer. A non-zero value for the 
quantities
\be
\frac{\ave{Z^k}}{\ave{N}} - \bar{y^k} ~~~~~~~~~~(k>1)
\label{eq25}
\ee
with
\be
\nonumber
\bar{y^k} = \int d\vec{p} ~ y^k(\vec{p}) ~ f^{incl}(\vec{p}),
\ee
would then indicate the presence of strong many-body correlations and
non-statistical fluctuations in the system. A
non-vanishing value of, e.g., $\frac{\ave{Z^2}}{\ave{N}} - \bar{y^2}$ 
indicates
the presence of two-body correlations from one part, and of possible 
correlations
between the particle multiplicity and the momentum
distributions from the other part. The absence of correlations between the 
particle multiplicity and the momentum
distributions alone does not necessarily imply a zero value for this quantity
\cite{gazd1,gazd3}. The generalization of this method to different particle
distributions, as for instance hadronic type distributions \cite{gazd2}, is
straightforward. In this case,
the N-particle momentum distributions 
$f_N(N,\vec{p_1},\cdot\cdot\cdot,\vec{p_N})$
are replaced by the
N-particle hadronic-type distribution functions 
$f_N(N,h_1,\cdot\cdot\cdot,h_N)$, where $h_i$ indicates the hadronic-type of
particle $i$, with the normalization
\be
\sum_{h_1=1}^{N_h} \cdot\cdot\cdot \sum_{h_N=1}^{N_h} 
f_N(N,h_1,\cdot\cdot\cdot,h_N) = P(N).
\ee
Here $N_h$ indicates the number of all possible hadron types. The event
observable is defined in this case as
\be
Z = \sum_{i=1}^{N} y(h_i),
\ee
where $y(h_i)$ is any function of the hadronic type of particle $i$.
The vanishing of the quantities in Eq.(\ref{eq25}) would indicate 
in this case an uncorrelated hadron-type production in the system.

A remark here is in order: There is a common misinterpretation of the quantity 
$\Phi = \sqrt{\frac{\ave{Z^2}}{\ave{N}}} - \sqrt{\bar{y^2}}$:
it has been argued 
that a
zero-value of this quantity would
indicate an equilibration (thermal or chemical) of the system
\cite{gazd2,blei2,gazd3,roland}. 
We have shown that such a
zero value of $\Phi$ merely indicates the absence of 2-body correlations and of
correlations of the particle multiplicity with momentum distributions.  No
specific form for the distribution function was assumed. 
It is true that a possible
equilibration of the system implies a completely uncorrelated system and 
gives a zero-value
for this quantity and all higher moments, but a zero value for 
$\Phi$ does not necessarily imply a complete equilibration of the system.

Note that the derivation of the quantities in Eq.(\ref{eq25}) was done for
a system of classical particles. For quantum systems, the s-body distribution
functions (Wigner functions) can not be written as a product of uncorrelated 
one-body distribution functions as in
Eq.(\ref{eq11}). It is clear that the only source of 
correlations in a classical system is the
existence of interactions between the particles. However, in a quantum system, 
there exists another
source of correlations: the existence of quantum-statistical boson or
fermion constraints. These are present even in an ideal gas of non-interacting
particles. Due to the presence of these minimal correlations coming from the 
quantum
nature (bosonic or fermionic) of the particles, the moments of the event 
observable $Z$ do not reduce for a
quantum system to the
moments of inclusive particle momentum distributions as in the classical case
(Eqs.(\ref{eq20}-\ref{eq22})), and the quantities in Eq.(\ref{eq25}) will not
vanish, even for a gas of independent particles \cite{mrow2}.

In conclusion, we have investigated the moments of event observable
distributions for systems with non-fixed multiplicity. These moments are shown
to be related to the higher many-body correlations. In the absence of any
correlations, these moments reduce for classical systems to the moments of 
inclusive particle
momentum distribution. Moreover, we have shown that a non-zero value for the 
quantities defined in Eq.(\ref{eq25})
indicates the presence of many-body correlations and 
non-statistical fluctuations in the momentum distributions of multiparticle
processes. 
We have also shown that the vanishing of the $\Phi$-variable does
not necessarily indicate an equilibration (thermal or chemical) of the system.

This work was supported by BMBF, GSI; DFG and Graduiertenkolleg
"Schwerionenphysik". M. Bleicher was supported by the Josef Buchmann
Foundation.

%\end{multicols}

\end{document}